\documentclass[runningheads]{lncse}
\usepackage[]{graphicx}

\newcommand{\DIR}{.}

\begin{document}

\title{CA Models for Traffic Flow: Comparison with Empirical 
  Single-Vehicle Data}

\titlerunning{CA models for traffic flow}

\author{W. Knospe\inst{1} \and L. Santen\inst{2} \and
  A. Schadschneider\inst{2} \and M. Schreckenberg\inst{1}}
\authorrunning{W.~Knospe, L.~Santen, A.~Schadschneider, and M.~Schreckenberg}

\institute{Physik von Transport und Verkehr,
  Universit\"at Duisburg, Germany \and Institut f\"ur
  Theoretische Physik, Universit\"at zu K\"oln, Germany}

\maketitle

\begin{abstract}
  Although traffic simulations with cellular-automata models give meaningful
  results compared with empirical data, highway traffic requires a more
  detailed 
  description of the elementary dynamics. Based on recent
  empirical results we present a modified Nagel-Schreckenberg cellular
  automaton model which incorporates both a slow-to-start and an anticipation
  rule, which takes into account  especially brake lights. The focus in this
  article lies on the comparison with empirical single-vehicle data.
\end{abstract}

\section{Introduction}

For a long time the modelling of traffic flow phenomena was dominated by two
theoretical concepts (for a review, see e.g.\ \cite{review}): 
Microscopic car-following models
and macroscopic models based on the analogy between traffic flow 
and the dynamics of compressible viscous fluids. 
Both approaches are still used widely by traffic engineers but for practical
purposes they are often not suitable, e.g. an efficient 
implementation for computer simulations of large networks is not possible. 
Macroscopic models use a large number of parameters which have partly 
no counterpart within empirical investigations. 
Moreover, the information which can be obtained using macroscopic models
is incomplete in the sense that quantities concerning individual cars 
cannot be introduced or derived directly.

In order to fill this gap cellular automata (CA) models have been invented
\cite{nagel93}. CA's 
are microscopic models which are by design well suited for large-scale
computer simulations. A comparison of the simulations with empirical
data shows that already very simple approaches give meaningful results. In
particular they can be used to simulate dense networks like
cities~\cite{esser}  which
are controlled by the dynamics at the intersections. However, for highway
traffic a more detailed description of the dynamics seems to be necessary. 

Recent empirical results show the existence of metastable states in
traffic dynamics and the occurrence of synchronised
flow~\cite{kerner_prl81,kerner_prl79,kerner96}, which can be identified by
vanishing cross-correlations of the local density and the local
flow~\cite{neubert}.  
Moreover, a detailed analysis of single-vehicle data~\cite{neubert} revealed
important facts for the microscopic modelling of traffic.
The time-headway distribution shows two characteristic peaks.
Small time-headways ($\approx 0.8~sec$) are a result of cars or
clusters of cars moving with small headway but large velocity, a 
time-headway of $2~sec$ can be identified with the drivers efforts for 
safety: It is recommended to drive with a distance of $2$ seconds.
Additionally, the distance-headway gives the most important information for
the adjustment of the car's speed for the correct description of the car-car
interaction. This is introduced in several models by the so-called optimal
velocity 
(OV) curve. It has been shown that one universal optimal velocity curve for all
density regimes does not exist, but individual curves for different densities
can be 
calculated (see the article of Neubert {\it et al.}\ in these proceedings for
more 
details). 
In fact, some model extensions of the CA model proposed by Nagel
and Schreckenberg (NaSch)~\cite{nagel93} exist which are capable to reproduce
metastable states~\cite{robert} or small 
time-headways~\cite{wolfgang,barret}, but
up to now it it not possible to generate synchronised traffic and
the correct microscopic properties mentioned above.

Here we propose a new CA model generalising the NaSch model and
some earlier extensions.
We compared our simulations with the corresponding data used
in~\cite{neubert}. 
The simulation data are evaluated by an artificial counting loop,
i.e. we measured the speed and the time-headway of the vehicles at a given
link of the lattice. This data set is analysed using the  
methods suggested in \cite{neubert}. 
In particular the density is calculated via the relation $\rho = J/v$ where
$J$ and $v$ are the mean flow and the mean velocity of cars passing the
detector in a time interval of $1$ minute. 
This dynamical estimate of the density gives correct results only
if the velocity of the cars between two measurements is constant, but
for accelerating or braking cars, e.g. in stop-and-go traffic, the
results do not coincide with the real occupation.  
In addition to the aggregated data also the single-vehicle data of
each passing car are analysed.
Although the empirical data have been obtained on a two-lane highway, the
simulations are performed on a single-lane road with one type of cars, because
the empirical results show no 
systematic lane dependence which is a consequence of the applied speed
limit.
 
In Sect.~\ref{new} we give a brief description of the new model definition
consisting of the NaSch-rules and some extensions. 
Section~\ref{validation} compares the simulation results of the new model with
the 
corresponding empirical data. Finally, Sect.~\ref{discussion} concludes with
a short summary and discussion. 

\section{New Approach}
\label{new}

Traffic networks can be classified as complex systems 
of a multitude of individual interacting agents.
In contrast to urban traffic where the flow is dominated by
intersections, signals etc., car characteristics like different
maximal velocities, acceleration capabilities and car lengths become
important on highways. In order to allow for a more realistic 
modelling of these characteristics we reduce the cell length of the
standard NaSch model (see~\cite{nagel93} for a detailed description of the
model) to a length of $l=1.5~m$. The acceleration and 
randomisation remains unaltered with one site per time step of $1~sec$
which leads to a velocity discretisation of $5.4~km/h$ which is slightly 
above "comfortable" acceleration of about $1~m/sec^{2}$~\cite{ite}.

The update rules of the new model combine the original
NaSch model and some recent extensions, namely a slow-to-start
rule~\cite{robert} and an anticipation term~\cite{wolfgang,barret}.
The slow-to-start rule allows to tune the velocity of the upstream
front of a traffic jam directly. It turns out that for a realistic 
choice of the parameters the outflow of a jam does not achieve the
capacity of the road. This empirical fact is known to lead to the 
existence of metastable states.

The next step towards a more realistic description of especially highway
traffic  
is to introduce anticipation effects, i.e. the adjustment of speed
also takes into account the expected behaviour of the leading
vehicle. Anticipation leads to a much more efficient lane-usage in
multi-lane traffic. Although both modifications significantly increase
the realism of the simulations a complete description of traffic 
highway traffic is not yet possible. The main problem with the existing
discrete models is that they fail to reproduce
platoons of slow moving vehicles. These patterns are not as stable as
in real traffic, i.e. the models overestimate the probability to form 
large compact jams.

This deficiency motivated us to prolong the range of interactions if a
braking maneuver of the leading vehicle happens or, more figurative,
we equipped the vehicles with brake lights. The event driven
interaction leads to a timely adjustment of speeds and therefore to a
more coherent movement of the vehicles in dense traffic. We
implemented the reaction to a brake light simply by an increased randomisation
parameter  $p_{b} > p $.

\section{Validation of the Model}
\label{validation}

\begin{figure}[hbt]
\begin{center}
 \includegraphics[width=.49\textwidth]{\DIR/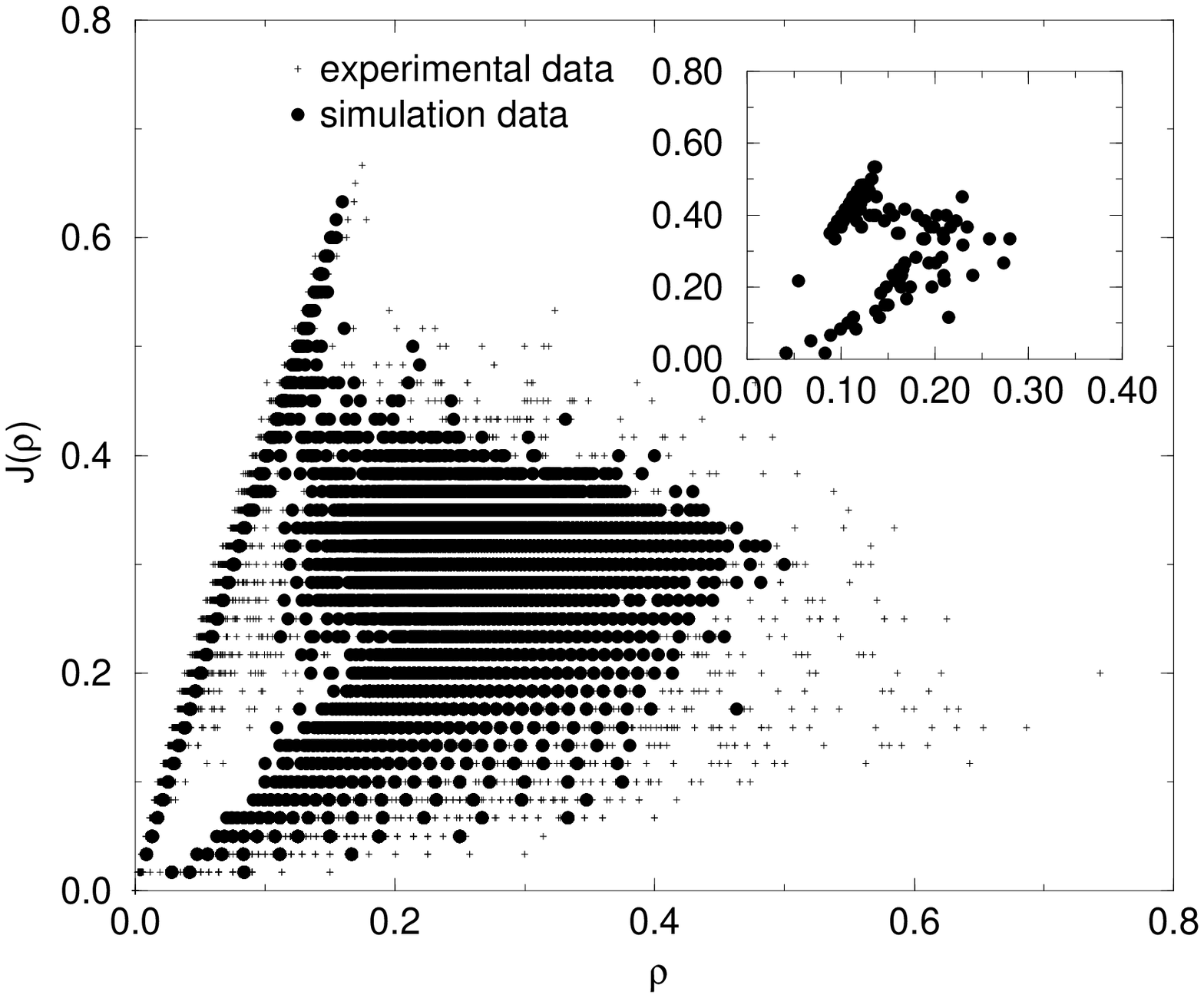}
 \includegraphics[width=.49\textwidth]{\DIR/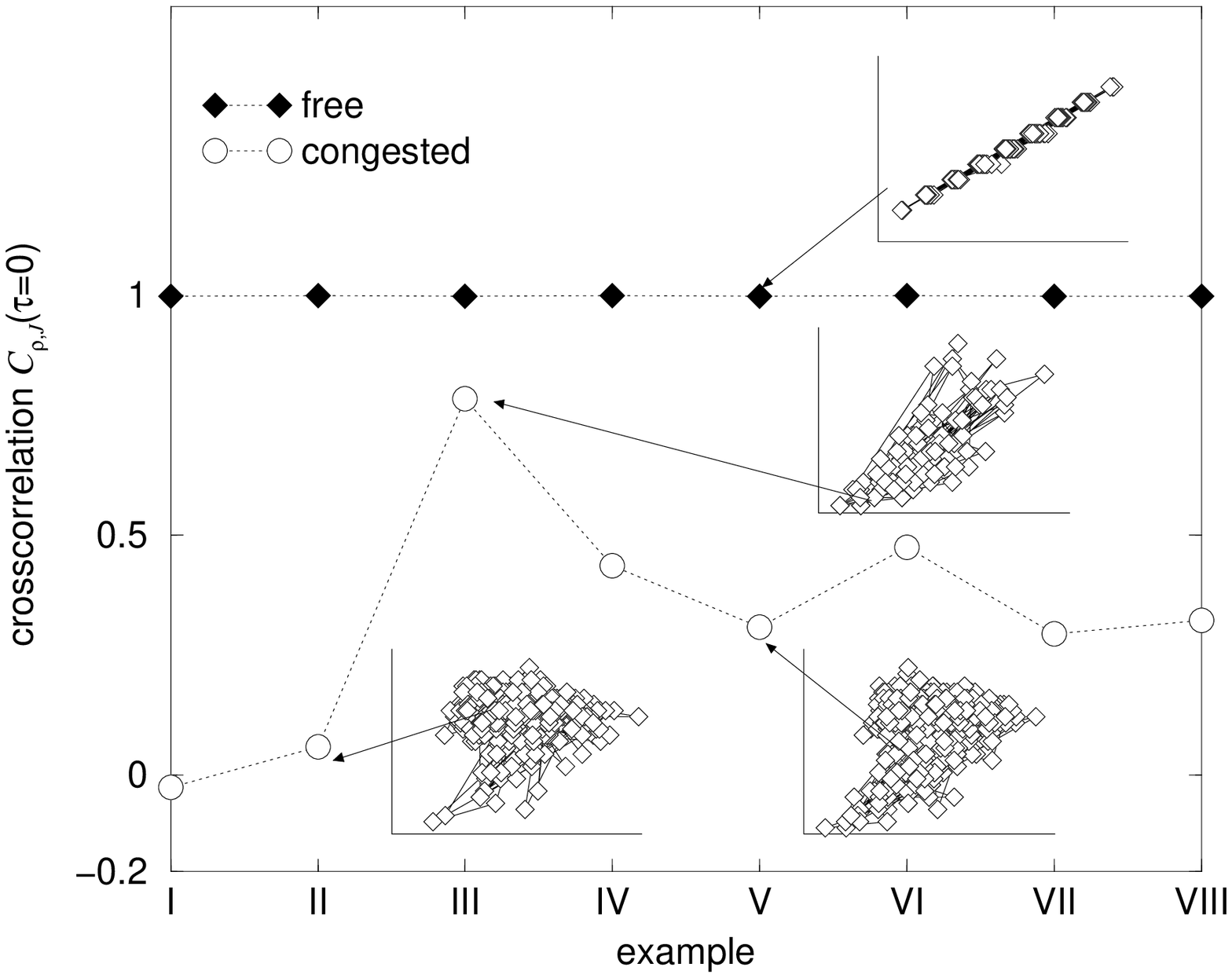}
\caption{Left: Comparison of the local fundamental diagram obtained by a
  simulation with the corresponding empirical fundamental diagram
  of Neubert {\it et al.} The inset shows the outflow of a megajam. Right:
  Cross-covariance of the flow and the density for  
  different global densities and homogeneous initialisation.}
\label{carnaschlocal}
\label{crosscovariance}
\end{center}
\end{figure}

Obviously, the fundamental diagram of the new model coincides very well
with the empirical data (Fig.~\ref{carnaschlocal}).
In comparison we observe a more narrow
distribution of densities. This further narrowing is simply an artifact 
of the discretisation of the velocities which
determines the upper limit of detectable densities.

The slow-to-start rule has been introduced in order to reduce the outflow of a
jam.   
This rule is responsible for the formation of large jams at high densities. 
To measure the outflow we used a megajam initialisation.
Obviously, the outflow is reduced considerably 
(inset of Fig.~\ref{carnaschlocal}). 
Using the autocorrelation function of the measured local density of a
system initialised with 
a megajam it is possible to determine the jam velocity. A detailed 
analysis leads to a value of about $12.75~km/h$.

As a next step for the validation of the model we compared single-vehicle
data of the simulation with the corresponding empirical data.
In order to give a correct comparison of our simulation data with the data of
Neubert {\it et al.} we tried to identify the three traffic states 
found in \cite{neubert}.
We therefore analysed the local data by means of the average velocity. A
contiguous time series of minute averages above 
$25~m/sec$ was classified as free flow, otherwise as congested flow.
In Fig.~\ref{crosscovariance} the cross-covariance $cc(J,\rho)$
of the flow and the local measured density for different traffic states
is also shown. 
In the free-flow regime the flow is strongly coupled to the density 
indicating that the average velocity is nearly constant.
Also for large densities, in the stop-and-go regime, the flow is mainly
controlled by density fluctuations. In the mean density region there is a 
transition between these two regimes.
At cross-covariances in the vicinity of zero the fundamental diagram shows a
plateau which supports the interpretation of \cite{neubert} that 
synchronised flow leads to $cc(J,\rho) \approx 0$. In the further comparison 
of our simulation with the corresponding empirical data we  
used these traffic states for synchronised flow data and congested states 
with $cc(J,\rho) > 0.7$ for stop-and-go data.

\begin{figure}[hbt]
\begin{center}
\includegraphics[width=.49\textwidth]{\DIR/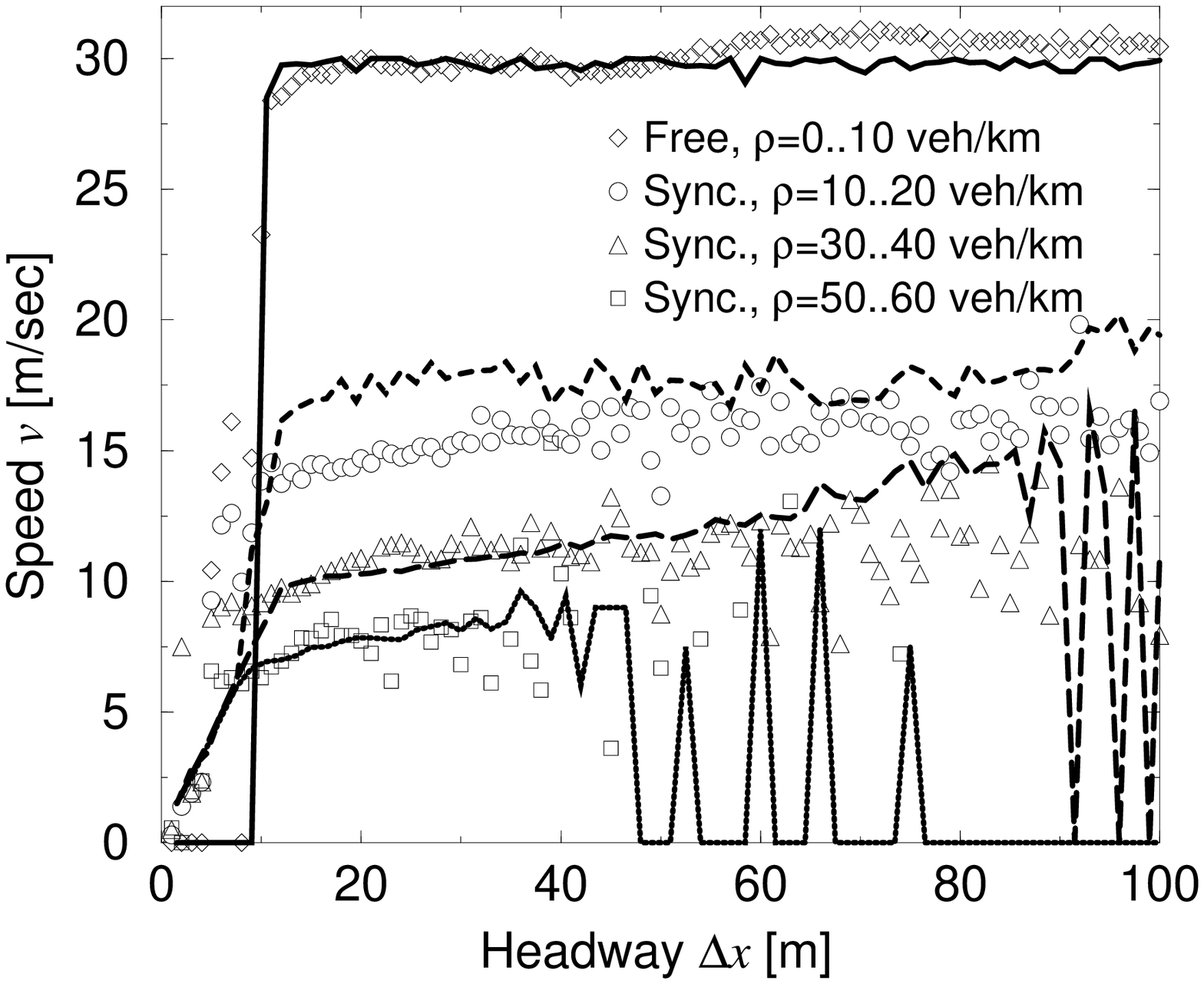}
\includegraphics[width=.49\textwidth]{\DIR/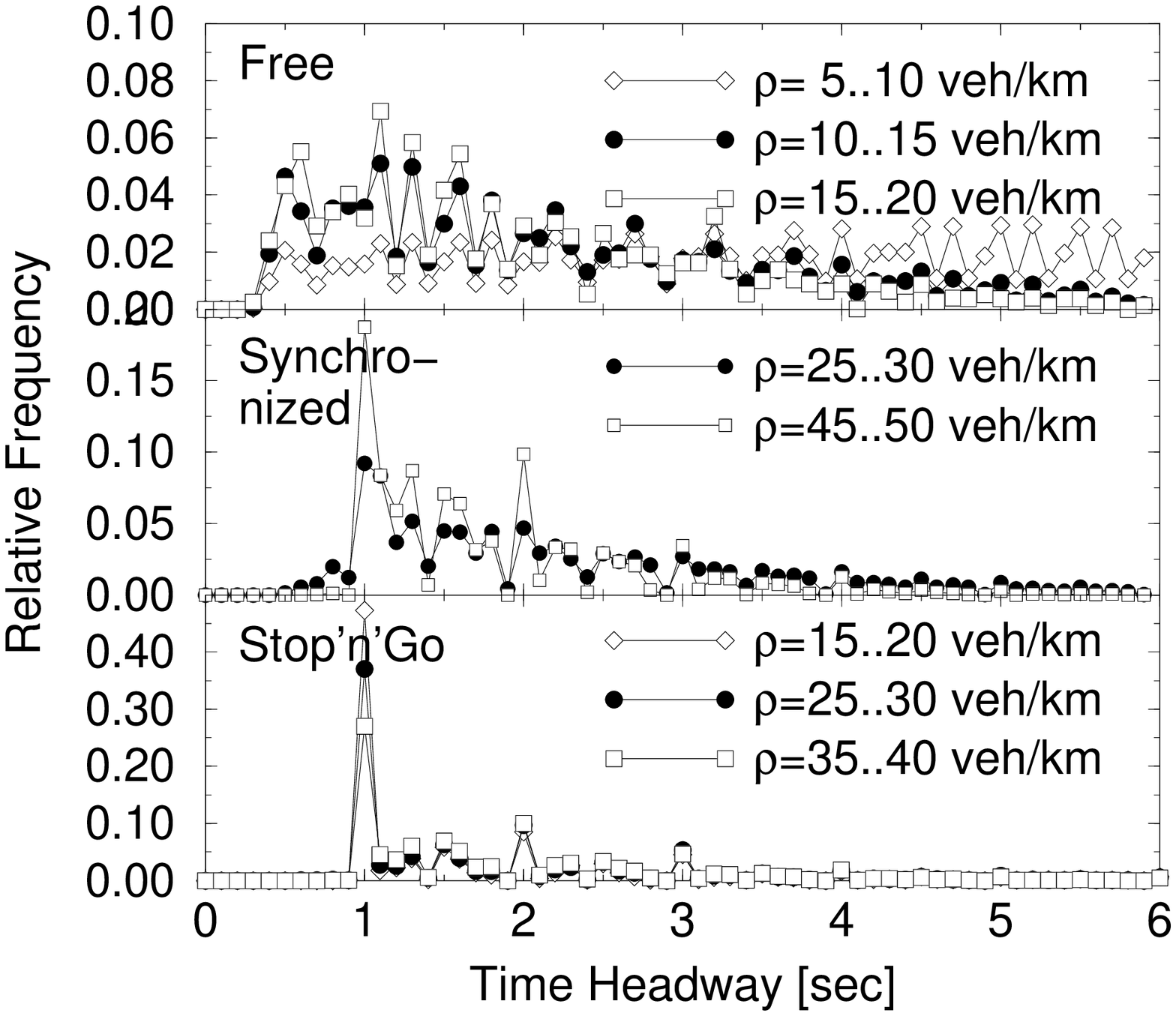}
\caption{Left: The mean speed chosen by the driver as a function of the gap
  to his predecessor. Comparison of 
  simulations (lines) with empirical data (symbols) of Neubert {\it et al.}
  Right: Time-headway distribution for different density regimes.}
\label{carnaschov}
\label{carnaschtime}
\end{center}
\end{figure}

For the correct description of the car-car interaction the distance-headway
(OV-curve) gives the most important information for the adjustment of 
the velocities. 
For densities in the free-flow regime it is obvious that the OV-curve
(Figure~\ref{carnaschov})
deviates from the linear velocity-headway curve of the NaSch model.
Due to anticipation effects smaller distances occur, so that driving with
$v_{max}$ is possible even within very small headways.
This strong anticipation becomes weaker with increasing density 
and cars tend to have smaller velocities than the 
headway allows so that the OV-curve saturates for large distances.
At headways of about $50~m$ the simulation data are in good agreement with
the empirical ones, but for large distances the acceleration behaviour of
the NaSch model cannot be suppressed so that the velocity increases with the
headway. 
In Fig.~\ref{carnaschtime} the time-headway distributions for different
density regimes are shown. The time-headways are calculated via the relation
$ \Delta t = {\Delta x}/{v} $ with a resolution of $0.1~sec$.
Due to the discrete nature of the model large fluctuations occur. In the 
free-flow state the anticipation rule is responsible for time-headways 
smaller than $1~sec$. The ability to anticipate the predecessors 
behaviour is getting weaker with increasing density so that small 
time-headways are nearly vanished in the synchronised and stop-and-go state.
Two peaks arise in these states: The peak at a time of $2~sec$ can be
identified with the driver's efforts for 
safety: It is recommended to drive with a distance of $2$ sec. 
Nevertheless, with increasing density the NaSch peak at a time of $1~sec$ (in
the NaSch model the minimal time-headway is restricted to $1~sec$) becomes
dominant. 
The higher the density the stronger this peak structure is pronounced.

\section{Summary and Discussion}
\label{discussion}

Based on empirical data we tried to find a simple extension of the original
NaSch model which is able to reproduce metastable states and synchronised
flow as well as microscopic features like density-dependent OV-curves and
characteristic time-headways.

First of all, the original NaSch cell length had to be reduced for a more
realistic acceleration behaviour which is especially important on
highways. For a more realistic car-car interaction anticipation terms seem to
play a crucial role. On the one hand, anticipation of the predecessors
movement in the next time step allows small time-headways and therefore high
flows. On the other hand, braking anticipation by means of brake lights
enables a driver to anticipate an imminent velocity reduction due to a jam. 
It is this braking anticipation which leads to synchronised states and to
increased time-headways which results in plateaus in the local fundamental
diagram.

Unfortunately, in this short contribution it is not possible to
describe all the features of the new model \cite{prep}. For example, 
a finer discretisation of the NaSch model or the corresponding limit 
$v_{max} \rightarrow \infty$ leads to metastable states analogous to 
the VDR-model which can be characterised by an order parameter.

For a further validation of this approach it is
necessary to extend the model to multi-lane traffic. The implementation of the
new model in the Online-Simulation of the Autobahn network in
Nordrhein-Westfalen should 
show its suitability for realistic traffic simulations.

\end{document}